\documentclass[aps,floatfix,prd,nofootinbib,superscriptaddress,reprint,showpacs,10pt,preprintnumbers,longbibliography]{revtex4-1}
\usepackage[utf8]{inputenc}
\usepackage[pdftex]{graphicx}
\usepackage{float}
\usepackage{amsmath}
\usepackage{amssymb}
\usepackage{mathtools}
\usepackage{siunitx}
\usepackage{amsfonts}
\usepackage{dsfont}
\usepackage{array}
\usepackage{bm}
\usepackage{mathrsfs}
\usepackage{pifont}
\usepackage{multirow}
\usepackage{upgreek}
\usepackage[dvipsnames]{xcolor}
\usepackage[pdftex,
  pdftitle={Put PDF title here},
  pdfauthor={Put PDF authors here},
  bookmarks,
  colorlinks,
  linkcolor=myblue,
  citecolor=mymagenta,
  menucolor=black,
  urlcolor=myblue,
  plainpages=false,
  pdfpagelabels,
  hypertexnames=false]{hyperref}
\usepackage{verbatim}
\usepackage{slashed}
\usepackage{cleveref}
\usepackage{ucs}
\usepackage{subfigure}
\usepackage{braket}
\usepackage[export=true]{adjustbox}
\usepackage{bbold}

\DeclareMathOperator{\im}{i}

\definecolor{mymagenta}{RGB}{200, 0, 100}
\definecolor{myblue}{RGB}{45, 48, 146}

\graphicspath{{plots/}}

\begin{document}
\title{Scaling and Lüscher Term in a non-Abelian (2+1)d SU(2) Quantum Link Model}
\author{Paul Ludwig}
\affiliation{Helmholtz-Institut f\"ur Strahlen- und Kernphysik, University of Bonn, Nussallee 14-16, 53115 Bonn, Germany}
\affiliation{Bethe Center for Theoretical Physics, University of Bonn, Nussallee 12, 53115 Bonn, Germany}
\author{Timo Jakobs}
\affiliation{Helmholtz-Institut f\"ur Strahlen- und Kernphysik, University of Bonn, Nussallee 14-16, 53115 Bonn, Germany}
\affiliation{Bethe Center for Theoretical Physics, University of Bonn, Nussallee 12, 53115 Bonn, Germany}
\author{Carsten Urbach}
\affiliation{Helmholtz-Institut f\"ur Strahlen- und Kernphysik, University of Bonn, Nussallee 14-16, 53115 Bonn, Germany}
\affiliation{Bethe Center for Theoretical Physics, University of Bonn, Nussallee 12, 53115 Bonn, Germany}
\date{\today}
\begin{abstract}
  We investigate a non-Abelian SU$(2)$ quantum link model in $2+1$
  dimensions on a hexagonal lattice using tensor network
  methods. We determine the static quark potential for a wide range of
  bare coupling values and find that the theory is confining. We also
  probe the existence of a Lüscher term and find a clear signal with a
  $g^2$ dependent coefficient, in qualitative agreement with a strong
  coupling expansion. Correspondingly, the width of the strings scales
  logarithmically with the string length again for all $g^2$-values,
  providing evidence for a rough string, with no indication
  for a roughening transition.
\end{abstract}

\maketitle
\section{Introduction}

Quantum gauge field theories constitute the fundament of the standard
model (SM) of particle physics. Lattice gauge theories provide the
non-perturbative approach required to understand quantum
chromodynamics, the strongly interacting sector of the SM.

With the advent of quantum computing technology, the quest is on to
find the most suitable formulation to simulate lattice gauge
theories on such future devices. Simulations on quantum computers in
the Hamiltonian formalism would allow probing non-perturbative
properties of gauge theories so far not accessible, see for instance
Ref.~\cite{DiMeglio:2023nsa}. One major challenge is to find
digitization strategies which allow one to map the gauge theory to the
irrevocably finite resources, even on quantum computers, while
preserving local gauge symmetry. And this challenge only grows when
the continuum limit is approached.

Tensor Networks (TNs) provide an efficient way of studying such digitization
even in the absence of reliable large scale quantum hardware, at least
as long as the state of interest does not encompass too much
entanglement. Originally conceived for problems in the 
field of many-body physics~\cite{Fannes:1990ur}, TNs efficiently approximate large classes of physical
states by truncating the entanglement~\cite{Hastings:2007iok}. Together with their efficient
implementation of the density matrix renormalization group (DMRG) algorithm \cite{Ostlund:1995zz} and
the time-dependent variational principle (TDVP) time evolution \cite{Haegeman:2011zz,Haegeman:2015ezw}
they represent state-of-the-art numerical tools for both ground state computation
and real-time evolution~\cite{Schollwoeck:2010uqf}.

The maybe standard approach to the challenge of digitizing lattice
gauge theories is to apply a suitable truncation strategy~\cite{Zohar:2014qma,Unmuth-Yockey:2018xak,Haase:2020kaj,Hartung:2022hoz,Kadam:2022ipf,Jakobs:2023lpp,Jakobs:2025rvz,DAndrea:2023qnr,Kadam:2024ifg,Fontana:2024rux,Jakobs:2025zcv} in
the framework of the original Kogut-Susskind Hamiltonian~\cite{Kogut:1974ag}.

Quantum link models (QLMs) provide an alternative approach
sidestepping the truncation and digitization problem. QLMs 
inherently preserve gauge symmetry exactly amid possessing a finite
number of degrees of freedom per gauge link. Originally formulated by Horn in
1981~\cite{Horn:1981kk} and subsequently studied by Orland and Rohrlich under
the name of gauge magnets~\cite{Orland:1989st}, they were later realized to
be a useful alternative regularization technique for non-Abelian gauge
theories~\cite{Chandrasekharan:1996ih} and extended to also
cover QCD \cite{Brower:1997ha}.

For this, the target symmetry group's link algebra, SU$(2)\times\text{SU}(2)$  is embedded in a
larger symmetry group, for instance SO$(5)$, for which a finite-dimensional 
representation $N_r$ needs then to be chosen to determine the
number of degrees of freedom per gauge link. The target continuum
gauge theory is recovered either by dimensional reduction~\cite{Beard:1997ic, Wiese:2021djl} or
by taking the limit $N_r\to\infty$. Therefore, in order to prepare for
the eventual execution of one of these approaches, QLMs and their
numerical simulation in the Hamiltonian formalism need to be
understood in detail. However, QLMs with a certain finite dimensional
representation of the embedding group are also in themselves
interesting quantum theories with exact gauge symmetry, and, therefore
deserve investigation. Previous works for U$(1)$ QLMs in $1+1$ and $2+1$ dimensions
have been performed using tensor networks~\cite{Pichler:2015yqa,Magnifico:2020bqt,Cardarelli:2017qop,Huang:2018vss,Felser:2019xyv,Hashizume:2021qbb,Halimeh:2021ufh} and cluster algorithms~\cite{Banerjee:2013dda,PintoBarros:2024oph} with several experimental simulations proposed~\cite{Osborne:2023rzx, Hauke:2013jga, Yang:2016hjn}. For SU$(2)$
previous works have focused on the $4$-dimensional representation of the embedding SO$(5)$ algebra~\cite{Banerjee:2017tjn,Mezzacapo:2015bra}.
There have also been some simulations of truncated SU$(2)$ in $2+1$d \cite{Cataldi:2025cyo, Cataldi:2023xki} which, for small electric truncations, is very similar to this model.

In this paper we investigate an SU$(2)$ QLM with the $5$-dimensional
representation of SO$(5)$ in $2+1$ dimensions on a 
hexagonal lattice in the
Hamiltonian formalism using tensor network states. We study the static
quark-antiquark potential in a wide range of values of the coupling
parameter, confirming that the theory shows confinement for all values
of the coupling. We determine the string tension from the
potential, which can be
used to set the scale of the theory. We find that from large values of
the coupling to intermediate ones the theory behaves similarly to what
is expected from Wilson's lattice gauge theories. However, for too small coupling values
the lattice spacing appears to diverge again, suggesting the non-existence of a
continuum limit in this QLM in the traditional sense. In addition, we probe the existence of the
so-called Lüscher term in the potential, and find clear evidence for
it. However, the dimensionless prefactor $\gamma$ does not yield the
expected universal value: the value of $\gamma$ is depending on $g^2$,
which we find to be in qualitative agreement with the strong coupling
expansion to first order.

The width of the string appears to scale logarithmically
with its length, indicating a rough string for all investigated couplings.
In this respect this QLM behaves differently as compared to the
$\mathbb{Z}_2$ gauge theory investigated in Ref.~\cite{DiMarcantonio:2025cmf}.
Moreover, the roughening transition and/or the Lüscher term have been
studied recently in Refs.~\cite{Xu:2025lve,Bombieri:2026czf}.

The remainder of this manuscript is organized as follows: In the
following section, we review the theory of the SU$(2)$ QLM and
introduce the static potential. Thereafter, in \cref{sec:method} we
discuss the methods and algorithms used. Our results are presented in
\cref{sec:results}, followed by a discussion in
\cref{sec:discussion} and conclusions in \cref{sec:conc}.
 
\section{Theory}

This section serves to introduce the $\mathrm{SU}(2)$ quantum link model and construct
its ring-exchange Hamiltonian. The starting point is the
Kogut-Susskind Hamiltonian~\cite{Kogut:1974ag} for the Wilson theory
\begin{equation}
  \label{eq:KSH}
  H = g^{2}\sum_{x,i}\underbrace{\left(\vec{L}^2_{x,i}+\vec{R}^2_{x,i}\right)}_{H_{E_{x,i}}} + \frac{1}{g^2}\sum_{P}\underbrace{\text{Tr}\left(U_P+U_P^\dag\right)}_{H_{M_P}}\,,
\end{equation}
composed of electric and magnetic parts, $H_{E_{x,i}}$ and
$H_{M_P}$, respectively. The plaquette operator is defined
as the product of link variables around the smallest closed loop, the
plaquette. For a square lattice it reads  
\begin{equation}
  \label{eq:plaq}
  U_{P(x,i,j)} = U_i(x)\cdot U_j(x+\hat i)\cdot U_i^\dagger(x+\hat j) \cdot U_j^\dagger(x)\,.
\end{equation}
Here, $i,j$ index the spatial directions only and $\hat i$ is a unit
vector in direction $i$. The generalization to other lattice
geometries is straightforward. The link variables are $\mathrm{SU}(2)$ matrices $U_i(x)$
fulfilling 
\[
U^\dagger\cdot U = U\cdot U^\dagger = \mathds{1}\,,\quad \det(U) = 1\,.
\]
The canonical momenta or electric operators $L$ and $R$ commute with
the link variables as follows
\begin{equation}
  \label{eq:ULR}
  \begin{split}
    \left[L^a_{x,i},U_{y,j}\right] &= -\delta_{xy}\delta_{ij}\frac{\tau^a}{2}U_{y,j}\,,\\
    \left[R^a_{x,i},U_{y,j}\right] &= \delta_{xy}\delta_{ij}U_{y,j}\frac{\tau^a}{2}\,.
  \end{split}
\end{equation}
The $\tau^a$ are Pauli matrices acting in colour space. The physical
Hilbert space $\mathcal{H}$ of this system is further restricted by
Gauss' law 
\begin{equation}
  \label{eq:Gauss}
  G^a_x \ket{\psi} = \sum_i\left(L^a_{x,i}+R^a_{x-\hat{i},i}\right)
  \ket{\psi} = 0, \hspace{0.2cm}\forall \psi\in\mathcal{H}. 
\end{equation}
It should be noted that, to fulfil this construction the local Hilbert space must necessarily be
infinite-dimensional.
\subsection{The \texorpdfstring{$\mathrm{SU}(2)$}{Lg} Quantum Link
  Model}

QLMs are obtained by embedding the link-algebra of a Wilson theory
into the algebra of a larger Lie group.
In the case of $\mathrm{SU}(2)$ this means promoting the link
variables $U_{x,i}$ to so-called \textit{quantum link operators}
\begin{equation}
  U_{x,i} = U^0_{x,i}\mathds{1}+iU_{x,i}^a\tau^a, \hspace{0.5cm} U^\dag_{x,i} = U^0_{x,i}\mathds{1}-iU_{x,i}^a\tau^a.
\end{equation}
The embedding is then
performed by extending the usual commutation relations of the link algebra
\begin{align}
  \begin{split}
    \left[ L_{x,i}^a , L_{y,j}^b \right] &= \im \delta_{xy}\delta_{ij}\epsilon_{abc}L^c_{x,i}\,,\\
    \left[ R_{x,i}^a , R_{y,j}^b \right] &= \im \delta_{xy}\delta_{ij}\epsilon_{abc}R^c_{x,i}\,,\\
    \left[L_{x,i}^a , R_{y,j}^b\right]&=0,\\
    \left[L^a_{x,i},U^\dag_{y,j}\right] &= \delta_{xy}\delta_{ij}U^\dag_{y,j}\frac{\tau^a}{2}\,,\\
    \left[R^a_{x,i},U^\dag_{y,j}\right] &= -\delta_{xy}\delta_{ij}\frac{\tau^a}{2}U^\dag_{y,j}\,.
  \end{split}
\end{align}
with commutation relations for the link operators in addition to \cref{eq:ULR}
\begin{align}
  \begin{split}
    \left[U^0_{x,i},U^0_{y,j}\right]&=0,\\
    \left[U^0_{x,i},U^a_{y,j}\right]&= \im \delta_{xy}\delta_{ij}\left(R^a_{x,i}-L^a_{x,i}\right)\,,\\
    \left[U^a_{x,i},U^b_{y,j}\right]&= \im \delta_{xy}\delta_{ij}\epsilon_{abc}\left(R^c_{x,i}+L^c_{x,i}\right)\,,
  \end{split}
\end{align}
It should be noted that, since the elements of $U_{x,i}$ are here operators, the commutation relations for $U^\dag_{x,i}$ are no longer trivial.
By construction this then yields $\mathrm{SO}(5)$ with the $10$
generators $R^a_{x,i}$, $L^a_{x,i}$ and $U^\rho_{x,i}$. The Kogut-Susskind Hamiltonian \cref{eq:KSH} and Gauss' law \cref{eq:Gauss}
remain unchanged. 

For $\mathrm{SO}(5)$ the two lowest dimensional non-trivial representations are a $4$-dimensional spinor representation
and a $5$-dimensional vector representation. While the $\{4\}$ representation has many interesting properties that are examined in Ref.~\cite{Banerjee:2017tjn}, its link
states possess different $\mathrm{SU}(2)$ representations on each of
their ends marking the representation as decidedly non-Wilson-like. In this paper, the $\{5\}$ representation will be used due to its
natural similarity to the Wilson theory: it offers the advantage of containing only states that have the same $\mathrm{SU}(2)$ representation on either end of a link, a universal feature of Wilson theories.
It therefore has a localized flux and is, in fact, very similar to a low order character expansion of $\mathrm{SU}(2)$.

\subsection{The Rishon representation}
A convenient way to better visualize QLMs (and other theories) is the rishon representation. Here one introduces
two pairs of fermionic ladder operators for each link
\begin{equation}
  \left(c^{a\dag}_{x,i,S},c^{a}_{x,i,S}\right)\,.
\end{equation}
These fermions are known as rishons.
$c^{a\dag}_{x,i,S}$ creates a rishon of type $a\in \{1,2\}$ at the right ($+$) or left ($-$) end denoted by $S$ of the link $(x,i)$, while $c^{a}_{x,i,S}$ annihilates it.
They obey the standard anti-commutation relations:
\begin{align}
\begin{split}
    \left\{c^{a\dag}_{x,i,S_1},c^{b}_{y,j,S_2}\right\} &= \delta_{xy}\delta_{ij}\delta^{ab}\delta_{S_1S_2},\\
    \left\{c^{a\dag}_{x,i,S_1},c^{b\dag}_{y,j,S_2}\right\} &= \left\{c^{a}_{x,i,S_1},c^{b}_{y,j,S_2}\right\} = 0. 
\end{split}
\end{align}
These auxiliary operators are then used to re-express all previous operators. For $R$ and $L$ this yields
\begin{align}
  \begin{split}
    R^1_{x,i}&=\frac{c^{1\dag}_{x,i,+}c^{2}_{x,i,+}+c^{2\dag}_{x,i,+}c^{1}_{x,i,+}}{2}\,,\\
    R^2_{x,i}&=\frac{-ic^{1\dag}_{x,i,+}c^{2}_{x,i,+}+ic^{2\dag}_{x,i,+}c^{1}_{x,i,+}}{2}\,,\\
    R^3_{x,i}&=\frac{c^{1\dag}_{x,i,+}c^{1}_{x,i,+}-c^{2\dag}_{x,i,+}c^{2}_{x,i,+}}{2}\,,\\
    L^1_{x,i}&=\frac{c^{1\dag}_{x,i,-}c^{2}_{x,i,-}+c^{2\dag}_{x,i,-}c^{1}_{x,i,-}}{2}\,,\\
    L^2_{x,i}&=\frac{-ic^{1\dag}_{x,i,-}c^{2}_{x,i,-}+ic^{2\dag}_{x,i,-}c^{1}_{x,i,-}}{2}\,,\\
    L^3_{x,i}&=\frac{c^{1\dag}_{x,i,-}c^{1}_{x,i,-}-c^{2\dag}_{x,i,-}c^{2}_{x,i,-}}{2}\,.
  \end{split}
\end{align}
From $R^3$ and $L^3$ one may now identify the rishons of type 1 and 2
with colour charges of $+\frac{1}{2}$ and $-\frac{1}{2}$, pictorially
represented  as (\includegraphics[height=10pt,
  valign=c]{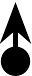}) and (\includegraphics[height=10pt,
  valign=c]{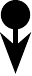}), respectively.
For link operators one obtains
\begin{align*}
  \begin{split}
    &U_{x,i}\\
    &=\begin{pmatrix*}
    \phantom{-}c^{2\dag}_{x,i,-}c^{2}_{x,i,+}+c^{1\dag}_{x,i,+}c^{1}_{x,i,-}& -c^{2\dag}_{x,i,-}c^{1}_{x,i,+}+c^{2\dag}_{x,i,+}c^{1}_{x,i,-}\\
    -c^{1\dag}_{x,i,-}c^{2}_{x,i,+}+c^{1\dag}_{x,i,+}c^{2}_{x,i,-}& \phantom{-}c^{1\dag}_{x,i,-}c^{1}_{x,i,+}+c^{2\dag}_{x,i,+}c^{2}_{x,i,-}
    \end{pmatrix*}\\
    &= \begin{pmatrix*}
      \phantom{-}\ \includegraphics[scale=0.3, valign=c]{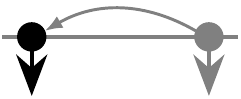}+\includegraphics[scale=0.3,valign=c]{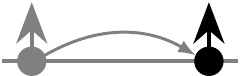}&-\ \includegraphics[scale=0.3, valign=c]{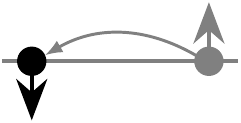}+\includegraphics[scale=0.3, valign=c]{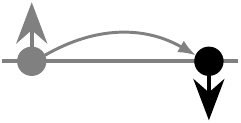}\\
      -\ \includegraphics[scale=0.3, valign=c]{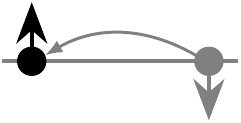}+\includegraphics[scale=0.3, valign=c]{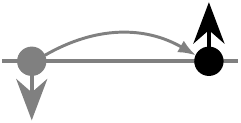}&\phantom{-}\ \includegraphics[scale=0.3, valign=c]{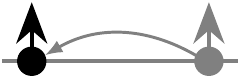}+\includegraphics[scale=0.3, valign=c]{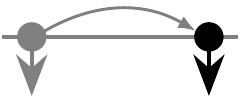}
    \end{pmatrix*}\\
  \end{split}
\end{align*}
\begin{align*}
  \begin{split}
    &U^{\dag}_{x,i}\\
    &=\begin{pmatrix*}
    \phantom{-}c^{2\dag}_{x,i,+}c^{2}_{x,i,-}+c^{1\dag}_{x,i,-}c^{1}_{x,i,+}& -c^{2\dag}_{x,i,+}c^{1}_{x,i,-}+c^{2\dag}_{x,i,-}c^{1}_{x,i,+}\\
    -c^{1\dag}_{x,i,+}c^{2}_{x,i,-}+c^{1\dag}_{x,i,-}c^{2}_{x,i,+}& \phantom{-}c^{1\dag}_{x,i,+}c^{1}_{x,i,-}+c^{2\dag}_{x,i,-}c^{2}_{x,i,+}
    \end{pmatrix*}\\
    &= \begin{pmatrix*}
      \phantom{-}\ \includegraphics[scale=0.3, valign=c]{tikz/tikz-figure9.pdf}+\includegraphics[scale=0.3, valign=c]{tikz/tikz-figure2.pdf}&-\ \includegraphics[scale=0.3, valign=c]{tikz/tikz-figure7.pdf}+\includegraphics[scale=0.3, valign=c]{tikz/tikz-figure4.pdf}\\
      -\ \includegraphics[scale=0.3, valign=c]{tikz/tikz-figure8.pdf}+\includegraphics[scale=0.3, valign=c]{tikz/tikz-figure3.pdf}&\phantom{-}\ \includegraphics[scale=0.3, valign=c]{tikz/tikz-figure6.pdf}+\includegraphics[scale=0.3, valign=c]{tikz/tikz-figure5.pdf}
    \end{pmatrix*}\\
  \end{split}
\end{align*}
where in the pictorial representation the greyed out state is annihilated and the black state created.
Here, it becomes apparent that the link operator transports each rishon to the other side of its link with a colour flip for the off-diagonal elements.
It also becomes clear that each of the operators conserves the total number
of rishons on each link
\begin{equation}
    \mathcal{N}_{x,i} = \sum_{a=1}^N \left(c^{a\dag}_{x,i,+}c^a_{x,i,+} + c^{a\dag}_{x,i,-}c^a_{x,i,-}\right)\text{.}
\end{equation}
One can show that this conserved quantity is equivalent to the choice
of a specific $\mathrm{SO}(5)$ representation. In particular for the
$\{5\}$ representation one has $\braket{\mathcal{N}_{x,i}} = 2$.\\
Thus, basis states for one link in the $\{5\}$ representation
are 
\begin{gather*}
  \ket{1}=\includegraphics[scale=0.3, valign=c]{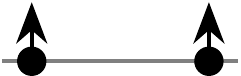}\,, \hspace{0.8cm} \ket{2}=\includegraphics[scale=0.3, valign=c]{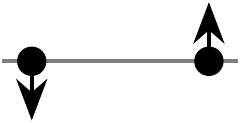}\,, \hspace{0.8cm} \ket{3}=\includegraphics[scale=0.3, valign=c]{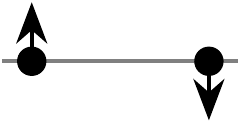}\,,\\
  \vspace{0.5cm}\\
  \ket{4}=\includegraphics[scale=0.3, valign=c]{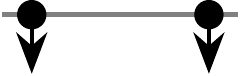}\,, \hspace{0.8cm} \ket{5}=\includegraphics[scale=0.3, valign=c]{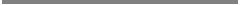}\,.
\end{gather*}
It should be noted that while $\ket{5}$ is drawn without rishons it is in fact the superposition of the two states where both rishons combine into a singlet at one end of the link.
\subsection{The Even Site Basis and the Ring-Exchange Hamiltonian}
We now work in a 2-dimensional hexagonal lattice.
The following construction was first introduced in \cite{Banerjee:2017tjn} for the $4$-dimensional representation.
We may now reduce the number of degrees of freedom of this model by utilizing local gauge invariance to suppress the type of rishon and uniformly depict them
as a dot ($\bullet$) thus reducing the possible states on a link from five to two. To recover the full physical Hilbert space,
one can now identify any configuration in the new basis that can fulfil Gauss' law
with a class of configurations in the original basis that are related by gauge transformations.
The configurations of the new basis that cannot fulfil Gauss' law in any type assignment
of rishons must still be removed. 
This can be done by further requiring that any site must hold an even
number of rishons. This leaves four options
\begin{center}
\begin{tabular}{ccccccc}
            \includegraphics[width=0.05\textwidth]{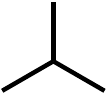} & \hspace{0.5cm} &
            \includegraphics[width=0.05\textwidth]{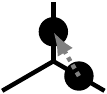} & \hspace{0.5cm} &
            \includegraphics[width=0.05\textwidth]{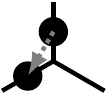} & \hspace{0.5cm} &
            \includegraphics[width=0.05\textwidth]{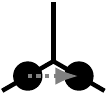}\\
            (1) && (2) && (3) && (4)
\end{tabular}
\end{center}
where the grey arrow indicates the sign of the singlet
state. Thereby the product state where the rishon at the tip of the arrow 
is identified as a $+\frac{1}{2}$ colour charge (\includegraphics[height=10pt,
  valign=c]{tikz/tikz-figure0.pdf}) carries the relative minus in the singlet.
Since each of the two remaining link states can be identified by just
one of its ends, it is sufficient to bipartition the hexagonal lattice
and construct a product basis from these four states at each even
(type A) site while imposing that an even number of rishons must meet
at each odd (type B) site as a new Gauss' law. This effectively
reduces the number of degrees of freedom to $\sqrt[3]{4} \approx 1.59$
per link and halves the number of Gauss' law equations.

\begin{figure}   
    \includegraphics[width=0.3\textwidth]{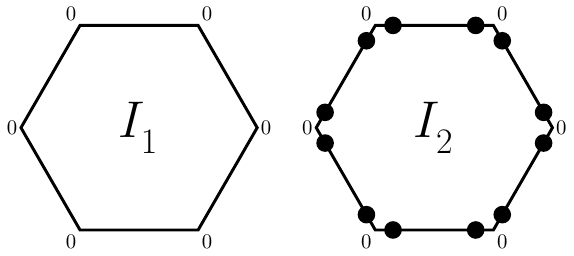}\\
    \vspace{0.25cm}
    \includegraphics[width=0.3\textwidth]{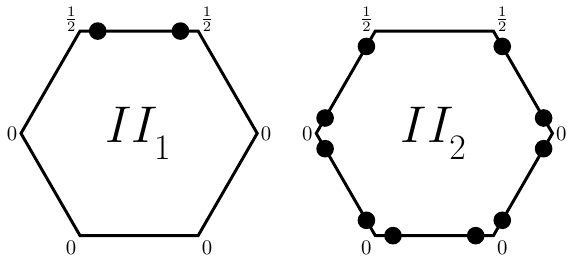}\\
    \vspace{0.25cm}
    \includegraphics[width=0.3\textwidth]{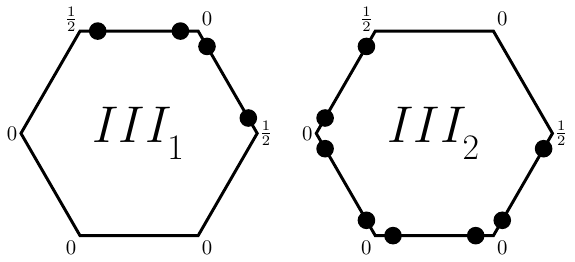}\\
    \vspace{0.25cm}
    \includegraphics[width=0.3\textwidth]{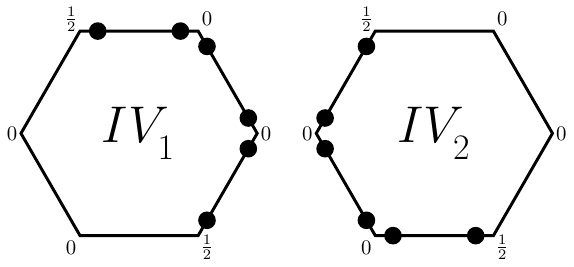}\\
    \vspace{0.25cm}
    \includegraphics[width=0.3\textwidth]{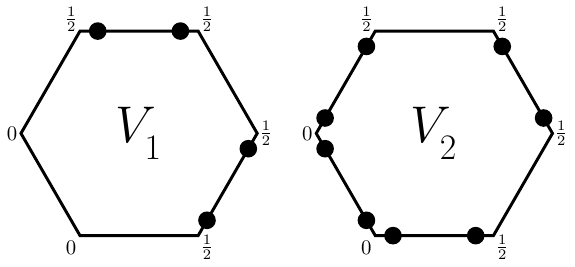}\\
    \vspace{0.25cm}
    \includegraphics[width=0.3\textwidth]{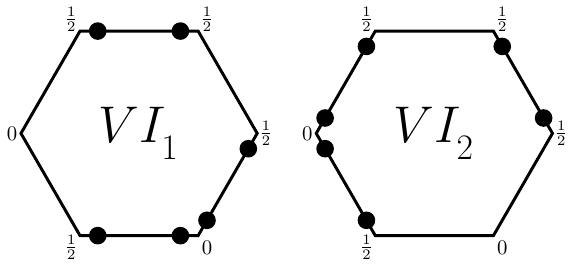}\\
    \vspace{0.25cm}
    \includegraphics[width=0.3\textwidth]{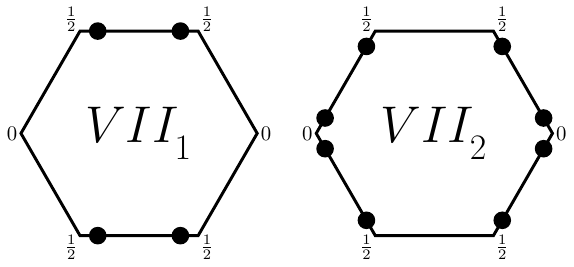}\\
    \vspace{0.25cm}
    \includegraphics[width=0.3\textwidth]{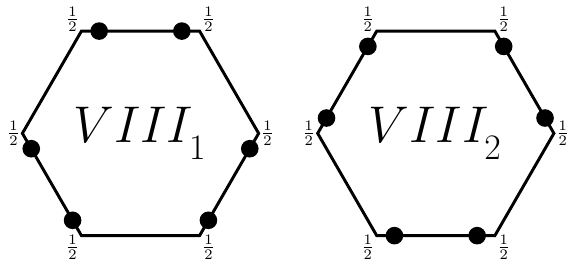}\\
    \caption{All eight possible plaquette environments for the $\{5\}$ representation, each with their two supported configurations. The fractions represent the charges external to the plaquette at the six sites.}
    \label{fig:envs}
\end{figure}

What remains is only to modify the Hamiltonian for this
construction. Since the electric term is already diagonal and
degenerate in the types of rishons, it requires no modification. 
The magnetic term can be rewritten in a general way by
realizing that the action of the plaquette operator cannot affect the
six links externally attached to the plaquette. 
Thus, any non-zero element of the Hamiltonian can only be between two
plaquette configurations with the same external link states which will
now be called \emph{environments}. 
Sorting all gauge invariant plaquette configurations one realized
that, up to rotation and mirror symmetries of the lattice, there are
only eight environments with two possible configurations each as can
be seen in \cref{fig:envs}. 

The Hamiltonian is thus block-diagonal along each of the eight environments which are specified by a hermitian $2 \times 2$ matrix
\begin{equation}
    H_\mathrm{env} = \begin{pmatrix*}
        W^1_\mathrm{env}&T_\mathrm{env}\\
        T_\mathrm{env}^*&W^2_\mathrm{env}
    \end{pmatrix*}.
\end{equation}
Additionally, for many of the environments the two supported configurations are related by either rotation or reflection requiring
\begin{equation}
    W^1_\mathrm{env} = W^2_\mathrm{env} = W_\mathrm{env}, \hspace{1cm} T_\mathrm{env} = T_\mathrm{env}^*.
\end{equation}
This yields a general Hamiltonian known as the ring-exchange Hamiltonian with 26 real parameters. For the QLM case
on the hexagonal lattice, the original magnetic Hamiltonian of the $\{5\}$ representation is recovered for $W=0$, $T_{I}=-128$, $T_{II}=T_{III}=T_{IV}=32$,
$T_{V}=T_{VI}=T_{VII}=-32$ and $T_{VIII}=32$ as can be shown by a
straightforward but lengthy reconstruction of the plaquette states in
terms of the original basis which shall be omitted here for brevity. 

\subsection{The Static Potential}
Since Gauss' law conserves the flux at each site, there can only exist physical states for which the amount of flux
entering and leaving the system on its edges is equal. Therefore, the case where at two locations external links transport $\pm \frac{1}{2}$ units of colour charge leads to the creation
of an unbreakable flux string between the two links. Thus, the model is confining. The mass of this string, meaning the energy of its ground state relative to the ground state without the string, depending on the string's length
is known as the static potential. The form of this potential in a $(2+1)$-dimensional Yang-Mills theory is known from effective string theory \cite{Luscher:2002qv} to be
\begin{equation}
  \label{eq:V}
  V(r) = \sigma r + \frac{\gamma}{r} + \mu + \mathcal{O}\left(\frac{1}{r^3}\right)
\end{equation}
at zero temperature. Here $r$ is the length of the string, $\mu$ is a regularization dependent mass and $\sigma$ is the string tension. The second term is known as the Lüscher term. It arises
from the zero-mode vibrations of the string. The value of $\gamma$ is
well known for Lorentz covariant systems to be $\gamma =
-\frac{\pi}{24}$ in two spatial dimensions \cite{deForcrand:1984wzs, Ambjorn:1984me, HariDass:2007tx, Caselle:2004er}. The Hamiltonian formulation
is not Lorentz-covariant, and thus a different value can arise as
argued in Ref.~\cite{DiMarcantonio:2025cmf}, where the authors also
observe $\gamma=-\pi/12$ instead. 

The Lüscher term only exists for rough strings and not for rigid
ones~\cite{Luscher:1980ac,Gliozzi:2010zv}. This is because in the rigid phase the
transverse excitations become gapped and thus do not contribute to the
ground state energy. These two types of strings can be differentiated by the
fact that for rough strings the squared width of the string $\omega^2$
increases logarithmically with its length, while for a rigid string it
is constant.

\subsection{Strong Coupling Expansion}

The strong coupling expansion provides a tool to gain analytical
insights for this QLM for large $g^2$-values, as has been also done
previously for Wilson's plaquette formulation, see for instance
Refs.~\cite{Itzykson:1980fz,Munster:1980ab}. That time, the roughening
transition has been estimated to occur at the $g^2$-value where the
string tension starts to deviate significantly from the leading order
strong coupling expansion~\cite{Munster:1980ab}. Moreover, it was
shown in Ref.~\cite{Kogut:1981ny} that if a diagonal string is studied
on the square lattice in Wilson's formulation, no roughening
transition occurs and $\gamma$ becomes $g^2$ dependent.

Here, we perform the strong coupling expansion of the relevant QLM to
leading, non-trivial order. To do so, we re-scale the Hamiltonian as
follows 
\begin{equation}
  \begin{split}
    g^{-2}H &=
    \sum_{x,i}\underbrace{\left(\vec{L}^2_{x,i}+\vec{R}^2_{x,i}\right)}_{H_{E_{x,i}}}
    +
    \frac{1}{g^4}\sum_{P}\underbrace{\text{Tr}\left(U_P+U_P^\dag\right)}_{H_{M_P}}\\
    &= H_E + \frac{1}{g^4} H_M
  \end{split}
\end{equation} 
and expand in $g^{-4}$. 
Starting with $0$th order, for two charges placed at a distance of
$R=2n$ with the geometry discussed in the previous sections, a basis
for the degenerate manifold $\mathcal{M}$ of minimal 
energy of $H_E$ is given by the $\binom{2n}{n}$ paths of minimal
length $4n$ between the charges. They can be enumerated by the strings
of length $2n$ containing $n$ letters $A$ and $n$ letters $B$. 
A particular string signifies that starting from the lower charge,
moving towards the upper charge the string at any branching point in
between chooses the right direction if the letter is $A$ and the left
one if the letter is $B$. The energy to $0$th order can then be read
off as
\begin{equation}
  \langle H_E\rangle^{(0)} = g^{-2}E^{(0)}= \left(\frac{3}{4} + \frac{3}{4}\right) 2 R\,,
\end{equation}
because all shortest paths have the same fixed length and contribute
the same flux.

To obtain the $1$st order terms we must first discuss the action of
the plaquette term on the basis states introduced above. To appear in
first order, the application of the plaquette term cannot change the
length of the string since otherwise it would leave $\mathcal{M}$. 
Since we know from previous sections that the action of the plaquette
activates all inactive links in a plaquette and deactivates all active
links, only applications of the plaquette term to plaquettes that are
adjacent to a charge string on exactly three sides contribute. This
corresponds to type VI in \cref{fig:envs}, for which we know the
matrix elements to be $32$.

This type of action is represented in the chosen basis by exchanging
two neighbouring letters $A$ and $B$. For this action we define the
operator $T_i$ 
\begin{equation}
    \begin{split}
        T_i \ket{\dots A_i B_{i+1} \dots} &= \ket{\dots B_i A_{i+1} \dots}\,,\\
        T_i \ket{\dots B_i A_{i+1} \dots} &= \ket{\dots A_i B_{i+1} \dots}\,,\\
        T_i \ket{\dots A_i A_{i+1} \dots} &= 0\,,\\
        T_i \ket{\dots B_i B_{i+1} \dots} &= 0\,.
    \end{split}
\end{equation}
Therefore, the $1$st order expansion can be written as
\begin{equation}
  g^{-2}H^{(1)} = g^{-2}E^{(0)} \mathds{1} + \frac{32}{g^4} \sum_{i=1}^{2n-1} T_i\,.
\end{equation}
If we identify $A$ with an occupation number of $1$ and $B$ with an
occupation number of $0$ (or the other way around), we may also write
\begin{equation}
  g^{-2}H^{(1)} = g^{-2}E^{(0)} \mathds{1} + \frac{32}{g^4}
  \sum_{i=1}^{2n-1} \left(c^\dag_i c_{i+1} + c^\dag_{i+1}
  c_i\right)\,. 
\end{equation}
Using a Jordan-Wigner map~\cite{Jordan:1928wi} one then obtains
\begin{equation}
  g^{-2}H^{(1)} = g^{-2}E^{(0)} \mathds{1} + \frac{32}{g^4}
  \sum_{i=1}^{2n-1} \left(\sigma^+_i \sigma^-_{i+1} + \sigma^-_{i}
  \sigma^+_{i+1}\right)\,, 
\end{equation}
with $\sum \sigma^z_i = 0$. Thereby, the second term on the right hand
side corresponds to an open XX chain at half filling. The ground state
energy of this model is known analytically~\cite{Bilstein:1999}, such
that we arrive at
\begin{equation}
    \begin{split}
    E^{(1)} &= 3 R\, g^2+ \frac{64}{g^2} \sum_{q=n+1}^{2n} \text{cos}\left(\frac{\pi q}{2n+1}\right)\\
    & = 3 R\, g^2 + \frac{32}{g^2} \left[1-\text{csc}\left(\frac{\pi}{2R+2}\right)\right]\,.
    \end{split}
\end{equation}
Expanding this expression for large $R$ yields
\begin{equation}
  E^{(1)} = \left(3 g^2  - \frac{64}{\pi g^2}\right)R +
  \frac{32-64/\pi}{g^2} - \frac{8\pi}{3 g^2} \frac{1}{R} +
  \mathcal{O}\left(R^{-2}\right)\,.
\end{equation}
Close to the strong coupling limit, therefore, one expects to find a
string tension of
$\sigma(g^2) = \left(3g^2 - \frac{64}{\pi g^2}\right)$
and a Lüscher term with coefficient
$\gamma(g^2)=- \frac{8\pi}{3 g^2}$. This result is analogous to the
one derived in Ref.~\cite{Kogut:1981ny} for diagonal strings on a
square lattice in Wilson's formulation.

Similarly, one can obtain the energy of the first 
excited state yielding a mass gap of
\begin{equation}
    M^{(1)} = \frac{128}{g^2} \text{cos}\left(\frac{\pi R}{2R+2}\right)
\end{equation}
again in the large $g^2$ limit.
 
\section{Methods}
\label{sec:method}

The primary method used in this paper is \textit{matrix product states} (MPS) which are then
used to determine the ground state with the density matrix renormalization group (DMRG) algorithm.
MPS approximates the wave function of a 1-dimensional lattice system by using singular value decomposition (SVD)
to split the tensor of the coefficient of the full state in the product basis into one smaller tensor for each site.
During this process singular values up to a certain combined size are neglected thus truncating the state efficiently. The number of remaining singular values $\chi$ is known as the \textit{bond dimension}.
The MPS ansatz also provides a highly efficient algorithm for evaluating similarly truncated \textit{matrix product operators} (MPOs)
on these states particularly for a local or almost local Hamiltonian.
This is then utilized by the DMRG algorithm which, beginning at a starting state minimizes the
energy of the state in terms of two neighbouring tensors per step while sweeping through the lattice.
In practice, it has proven efficient to start this algorithm at small bond dimensions in order to quickly converge
near the desired state and then successively increase the bond dimension to improve the approximation.
This method has the risk to converge to local minima. To mitigate this, the Hamiltonian is in some sweeps perturbed by a random noise term.
For more information on both MPS and DMRG the authors recommend \cite{Schollwoeck:2010uqf}.
In this paper MPS will be applied to two-dimensional lattices. This can be understood as transforming a
two-dimensional system of a nearest-neighbour Hamiltonian that is small in the second direction into a 1-dimensional system where
the Hamiltonian now has a range that is given by the original system's
extent in the second direction. While this slows the DMRG, the method
remains viable in particular for limited extents in the second
direction, as long as one can afford the higher bond dimensions 
required.

The Hamiltonian from the previous section is simulated in $(2+1)$D
using the DMRG implementation from the ITensor~\cite{Fishman:2020gel} library
for the Julia programming language~\cite{Bezanson:2014pyv}. 
Gauss' law is enforced by introducing a penalty term in the
Hamiltonian that lifts the unphysical states out of the low-energy
spectrum. From the $64$ possible configurations of the even site basis
around an odd site the Gauss' law breaking ones are selected and an additional
mass of $\kappa$ is added to them. $\kappa$ must then be tuned such that the expectation value
of the prohibited configurations vanishes in the result.

The bond dimension is dynamically chosen for each sweep limiting
the summed squares of the cutoff singular values to at most $10^{-9}$, with
the bond dimension initially being artificially limited and that
limit being raised every few sweeps. Additional noise terms in the MPS
are used for the first few sweeps. The DMRG is terminated when the
change in energy between sweeps has dropped below $10^{-5}$ with a
minimum number 
of sweeps being performed regardless to prevent a premature termination due to insufficient bond dimension. The DMRG code was compared to exact diagonalization
on a $4 \times 4$ basis site lattice and was
found to be in exact agreement.
The noise terms and the limits on the bond dimension were tuned for fast convergence and good agreement with the exact diagonalization of the $4 \times 4$ system with periodic boundaries. 
Since DMRG does not produce statistical
uncertainties, uncertainties on the fit parameters are computed from
the inverse Hessian of the $\chi^2$ function.

\subsection{Setup}

\begin{figure}
  \includegraphics[width=0.25\textwidth]{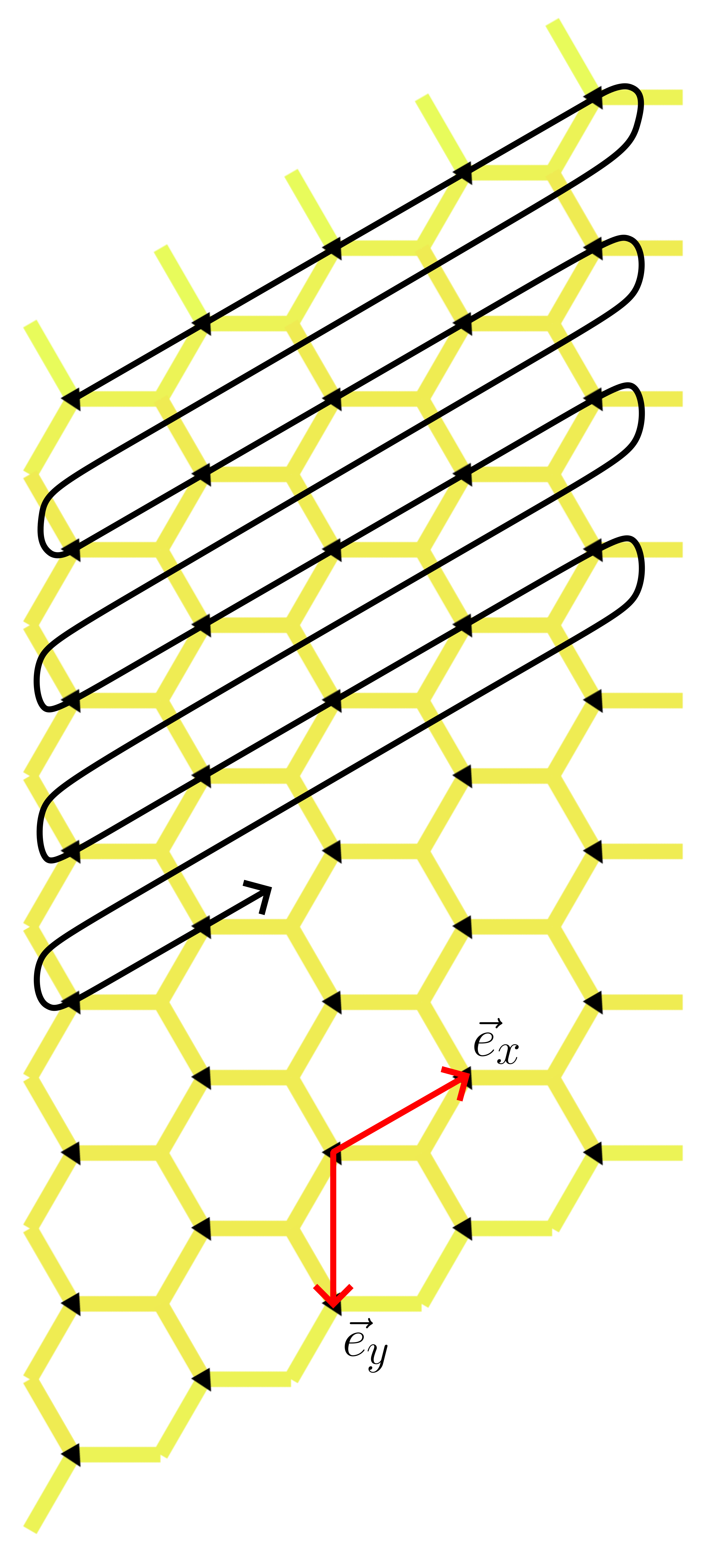}
  \caption{Exemplary plot of the simulated geometry for $N_x=5$ and $N_y=8$. The black triangles indicate the even site basis sites and the black
  arrow signifies the way in which the $1$d MPS is inlaid into the $2$d system.}
  \label{fig:setup}
\end{figure}

\begin{figure}
    \includegraphics[width=0.8\columnwidth]{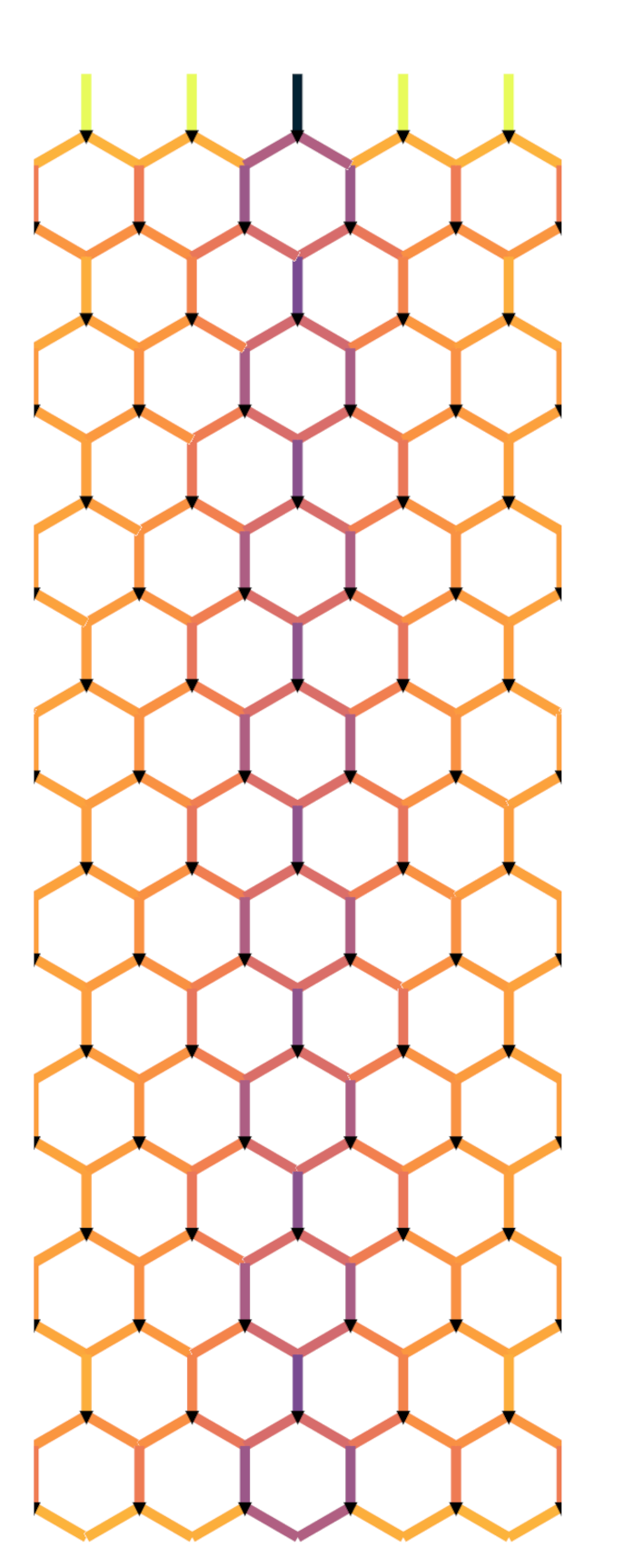}
    \caption{Exemplary plot of a flux string ground state at
      $g^2=1.5$. For each link, the expectation value of
      $\mathbb{1}-\ket{0}\bra{0}$ is plotted signifying the total
      flux carried by the link. Dark blue is $1$, yellow is $0$.} 
    \label{fig:string}
\end{figure}

The ground state is calculated for $N_x \times N_y$ systems (see \cref{fig:setup}) with
$N_x\in \{3,4,5\}$, periodic boundary conditions in the $x$-direction and closed\footnote{Closed meaning in this case that no flux may be transported into or out of the system on these boundaries.} boundary
conditions in the $y$-direction, both with and without an unbreakable
flux string in the $y$-direction. An example of this geometry can be seen in \cref{fig:setup} together with an example of a flux string in \cref{fig:string}. 
This was done for $N_y \in \{2,4,\dots,14,16\}$ and $g^2 \in \{0.5,0.75,\dots,7.75,8\}$. For a fixed system size, first the vacuum state, in case of the systems without $\frac{1}{2}$ charges on the edges, or otherwise a superposition of the vacuum with the shortest path string\footnote{The system with $\frac{1}{2}$ charges on the edges would not converge if the starting state did not contain
a closed string since the spontaneous creation of a string would require all contributing sites to be updated at the same time which DMRG does not and any incomplete string would be subject to the energy penalty from Gauss' law. The only alternative option would be to include very large noise terms for many sweeps and hope that the string appears from the noise, though, this is exponentially unlikely in the length of the string.} is used as an initial state
for $g^2=8$ which is fairly close to the true ground state
\cite{Kogut:1979wt}. The resulting ground state is then used as the
initial state for the DMRG at the next smallest $g^2$. For $g^2 = 8$,
the minimum sweeps are set to $50$ with $15$ noise sweeps to ensure
that the correct ground state is found. For all further calculations
the minimum sweeps are $7$ with $4$ of them noisy.
To prevent the MPS from exceeding the available RAM, a maximum bond dimension of $2000$ is imposed. However, based on the calculated truncation error, this does not seem to significantly affect the results. For the resulting states, the energy, the variance of the Hamiltonian, the expectation values of the penalty terms and the expectation values of the projectors onto each of the four even site basis states on each link are saved. Furthermore, for the chargeless systems only, the electric part of the energy is measured separately.

\subsection{Convergence}
\begin{figure}
  \includegraphics[width=0.45\textwidth]{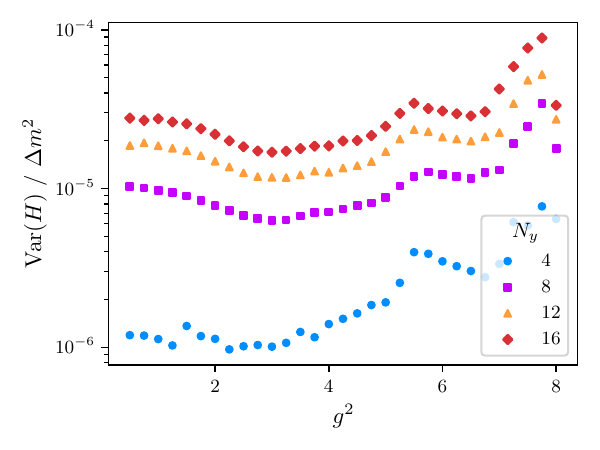}
  \includegraphics[width=0.45\textwidth]{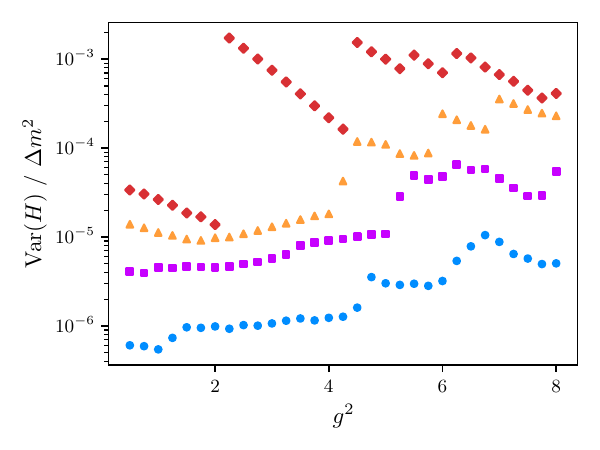}
  \caption{$\text{Var}(H)/(\Delta m)^2 \approx (1-f)$ for the empty systems (top) and the systems with strings (bottom).}
  \label{fig:conv}
\end{figure}

Since DMRG can get trapped in local minima, it is vital to check that convergence was achieved. For large systems the convergence to the true global
minimum is difficult to quantify. Here an approximation is used that states that for states that are close to the ground state
\begin{equation}
  \frac{\text{Var}(H)}{(\Delta m)^2} = (1-f) + \mathcal{O}\left((1-f)^2\right)
\end{equation}
holds where the variance is computed with respect to the DMRG output,
$\Delta m$ is the mass gap of the system and $f = |\braket{\Psi_\text{exact}|\Psi_\text{approx}}|^2$ is the fidelity.
Since the true mass gap is unknown it is substituted by the mass gap of the $4\times4$ lattice of the same topology obtained via exact diagonalization for all couplings. It should be noted that
the mass gap tends to slightly decrease as the volume increases, however by comparison to previous mass gap calculations \cite{Engels:1991ij} it can be
estimated that for the ratios of volumes required $(\Delta m)^2$ is at most overestimated by a factor of $10$. As can be seen in \cref{fig:conv} the
approximated infidelity is for many cases around or below $10^{-4}$ with some larger systems with strings around $10^{-3}$ indicating a
good convergence for all systems. The
infidelities both of larger systems and of systems containing strings are generally larger.

\begin{figure}
  \includegraphics[width=0.45\textwidth]{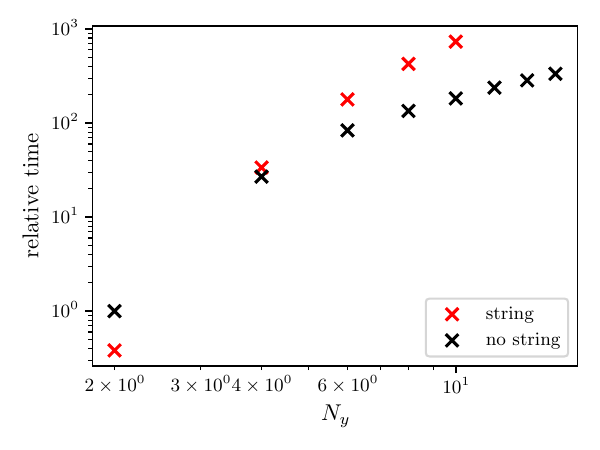}
  \caption{Total runtime for each coupling both with and without a
    string in a log-log plot for $N_x=5$.}
  \label{fig:runtime}
\end{figure}

As a side it should be noted that for the runtime of the DMRG two different behaviours were observed as plotted in \cref{fig:runtime}. For systems without a string
the runtime increased linearly starting from $N_y=6$ as would be expected of a system with a small correlation length.
For the systems with a string the runtime increased as a higher order polynomial. It seems likely that this higher order polynomial
is the effect of a large increase in the correlation length induced by the strings.

\section{Results}
\label{sec:results}

Using the methods described in the previous sections we have
investigated the QLM on the hexagonal lattice in a wide range of bare
couplings $0.5\leq g^2\leq 8$, with 8 additional points at
$95\leq g^2\leq 125$. Moreover, we study the dependence of the
potential $V$, \cref{eq:V} on the transversal
lattice extent $N_x$ with values $N_x=3,4$ and $N_x=5$.

\subsection{Electric versus Magnetic Contributions to $H$}

\begin{figure}
\includegraphics[width=0.95\columnwidth]{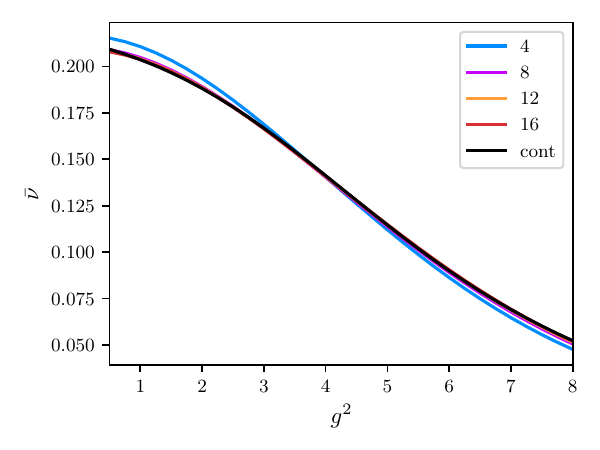}
  \caption{Fraction of flux transporting links $\bar\nu$ as a
    function of the coupling $g^2$ for different $N_y$ at $N_x$=5}
  \label{Fig:fluxlinks}
\end{figure}

It is well known that the Kogut-Susskind Hamiltonian exhibits a
magnetic regime at small $g^2$ where $H_M$ dominates and an electric
regime at large $g^2$ where $H_E$ dominates.
In this model they can be distinguished via the average fraction of
flux transporting links
\begin{equation}
  \bar\nu\ =\  \frac{1}{|\Lambda|} \sum_{i\in \Lambda} \langle 1-\ket{0} \bra{0} \rangle_i
\end{equation}
shown in \cref{Fig:fluxlinks}. Here $\Lambda$ is the set of all links
in the lattice. In the electric regime ($g^2 \to 
\infty$) this value goes to zero while it attains a finite value in
the magnetic regime, as can be seen in \cref{Fig:fluxlinks} where we
show $\bar\nu$ as a function of $g^2$ for different $N_y$.
Given the quick convergence to a universal curve for large $N_y$
visible in \cref{Fig:fluxlinks}, it seems likely that there is no
phase transition but a crossover between the two regimes. 

\begin{figure}
  \includegraphics[width=0.95\columnwidth]{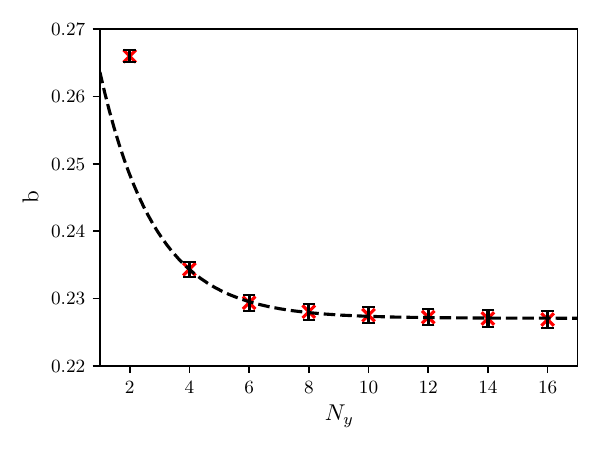}
  \caption{Infinite length extrapolation for the fraction of flux
    carrying links at $g^2=0$ with $\chi^2_{\text{red}} =
    0.03$ and $N_x=5$. The data point at $N_y=2$ is excluded from the fit. The parameters are $\alpha = -0.06(6)$, $\beta = 0.5(3)$ and $\gamma = 0.227(1)$. The errors are obtained from the inverse Hessians of the previous fits.}
  \label{Fig:ex_links}
\end{figure}

\begin{figure}
  \includegraphics[width=0.95\columnwidth]{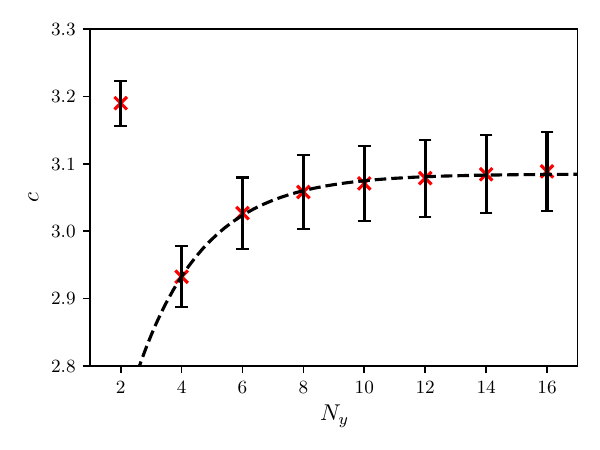}
  \caption{Infinite length extrapolation for the centre of the
    crossover with $\chi^2_{\text{red}} = 0.006$ and $N_x=5$. The data point at
    $N_y=2$ is excluded from the fit. The parameters are $\alpha = 1(2)$, $\beta = 0.5(5)$ and $\gamma = 3.09(4)$. The errors are obtained from the inverse Hessians of the previous fits.}
  \label{Fig:ex_crossover}
\end{figure}

Fitting this data with a generalized logistic function
\begin{equation}
  \bar\nu(g^2) = b(1+\exp(d(g^2-c)))^{-m}
\end{equation}
for each system length $N_y$, we can then perform the infinite length
extrapolations for the fraction of flux carrying links $b$  
(shown in \cref{Fig:ex_links}) and the centre of the crossover $c$ (shown in
\cref{Fig:ex_crossover}) using the ansatz
\begin{equation}
  \label{eq:exp}
  y = -\alpha\ \text{exp}(-\beta N_y)+\gamma\,.
\end{equation}
The fits can be performed assuming no correlation since the data points stem from independent simulations.
Excluding $N_y=2$ from the fits, we obtain the fraction at $g=0$ as 
$\bar\nu_{g=0} = 0.227(1)$ and the crossover point at $g^2 = 3.09(4)$
in the limit $N_y\to\infty$. 

\subsection{The static quark potential}

\begin{figure}
    \includegraphics[width=0.95\columnwidth]{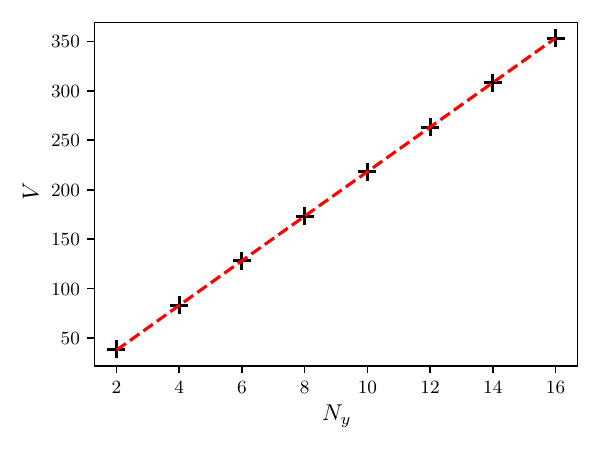}
    \caption{Potential $V$ as a function of $N_y$ for transversal
      lattice extent $N_x=5$ and
      $g^2=4$. The dashed (red) line represents a fit of \cref{eq:V} to the
      data for $N_y > 4$.}
    \label{fig:potfitexample}
\end{figure}

By subtracting the ground state energy of the system without charges
of the same size from the ground state energy of a system containing a
charge pair, one can compute the static quark potential $V(r)$
\cref{eq:V} for this theory. Since the charges are located at the upper and lower
boundaries, respectively, the potential $V$ is evaluated at distance
$r=N_y$, and we vary $N_y$ to probe different distances. The data for small $N_y$ is excluded from the fits as it contains large finite size and discretization effects. For $N_x=5$ only $N_y  \geq 6$, for $N_x=4$ only $N_y \geq 10$ and for $N_x=3$ only $N_y \geq 8$ are used.
The amount of the effect as well as the system lengths at which it occurs seems to depend both on the width of the system and on the parity of the system's width and length.
 We observe a
positive string tension $\sigma$ for all values of $g^2$ investigated,
indicating confinement. At least visually, there are barely deviations
from the linear dependence on $r$ visible. This is shown exemplarily
in \cref{fig:potfitexample}, where we show the potential in lattice
units as a function of $N_y \equiv \frac{2 r}{3}$ for $g^2=4$. Note
that the factor $2/3$ in the relation between $r$ and $N_y$ stems from
the hexagonal geometry in the investigated lattice.

\begin{figure}
    \includegraphics[width=0.95\columnwidth]{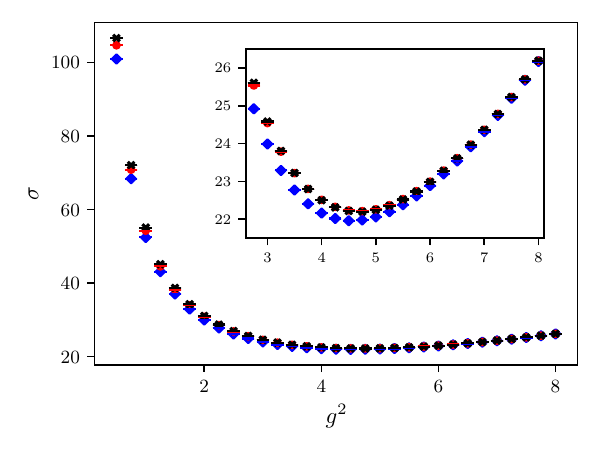}
    \caption{String tension $\sigma$ as a function of $g^2$ with $N_x=3$ (blue diamond), $N_x=4$ (red dot), and $N_x=5$ (black x).}
    \label{fig:tension}
\end{figure}

In \cref{fig:tension} we show the string tension $\sigma$ as a
function of $g^2$. The string tension measured in the simulations is
in so-called lattice units, i.e. $\sigma = a(g)^2
\sigma_{\mathrm{physical}}$, where $\sigma_{\mathrm{physical}}$ has
units of energy squared. The value of $\sigma_{\mathrm{physical}}$ is
not known, because there is no realization of this QLM in nature, but
its value in physical units is also not needed. If this universal
curve exists, one can use
$\sqrt{\sigma} = a\sqrt{\sigma_{\mathrm{physical}}}$ and its inverse
to quote distances and energies in units of (inverse)
$\sqrt{\sigma}$. If this theory indeed exhibits a universal observable
$\sigma_{\mathrm{physical}}$, $V(r\sqrt{\sigma})/\sqrt{\sigma}$ will
be a universal curve, up to discretization artefacts. This is the case
as can be observed in \cref{fig:uni}, with basically no lattice
artefacts visible.

\begin{figure}
    \includegraphics[width=0.95\columnwidth]{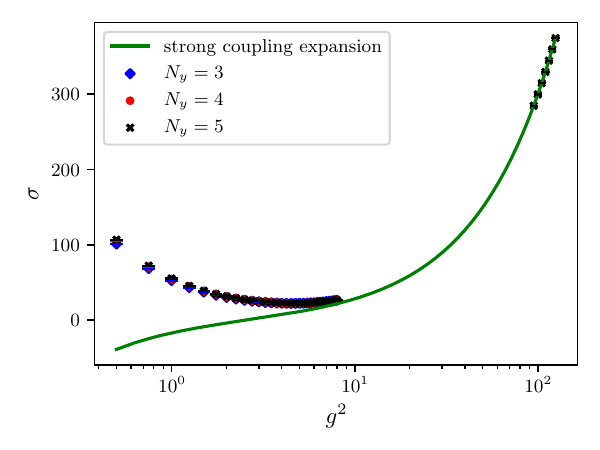}
    \caption{String tension $\sigma$ as a function of $g^2$ with $N_x=3$ (blue diamond), $N_x=4$ (red dot), and $N_x=5$ (black x) with the prediction from the strong coupling expansion added as a green line.}
    \label{fig:tension2}
\end{figure}

\begin{figure}
    \includegraphics[width=0.95\columnwidth]{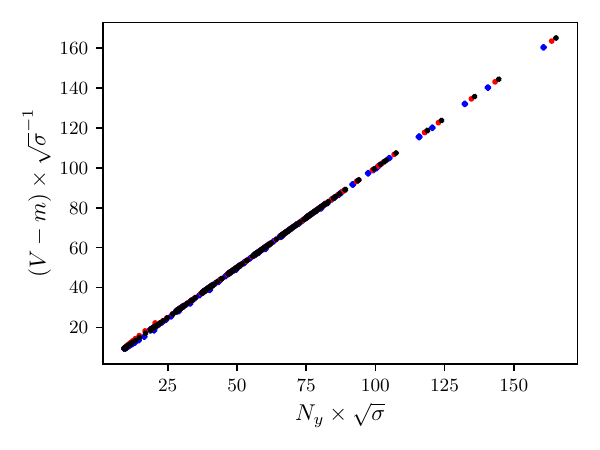}
    \caption{Static potential in units of $1/\sqrt{\sigma}$ as a
      function of the distance $N_y\sqrt{\sigma}$ including data for all
      values of $g^2$. The different colours represent data points with $N_x=3$ (blue diamond), $N_x=4$ (red dot), and $N_x=5$ (black x).}
    \label{fig:uni}
\end{figure}

The string tension, therefore, gives also access to the lattice spacing:
from \cref{fig:tension} it can be read off that for this model the
lattice spacing diverges both for strong and weak coupling. For $g^2=100$ the string tension was
additionally determined to be $\sigma = 299.79950(4)$ which is fairly consistent with the strong coupling expansion that predicts $\sigma = 299.7963$.
As can be seen in \cref{fig:tension2}, which shows the string tensions for different $N_x$ plotted over $g^2$ together with the prediction from the strong coupling expansion, similar agreement is found for all other data points at large couplings while, expectedly, the behaviour at intermediate and small coupling is not well predicted.
We remark that we have performed the equivalent procedure using the
so-called Sommer parameter~\cite{Sommer:1993ce} with the results being
in agreement within errors with the ones from the string tension. 
We also remark that in $2+1$ dimensions the coupling is
dimensionful, i.e.
\begin{equation}
  g^2 = a g_0^2
\end{equation}
with $g_0$ the bare running coupling of the theory. However, if one
used $1/a$ as the energy scale at which to define $g_0$, the bare
coupling $g_0(1/a)$ is not uniquely defined.

\begin{figure}
  \centering
  \includegraphics[width=0.95\columnwidth]{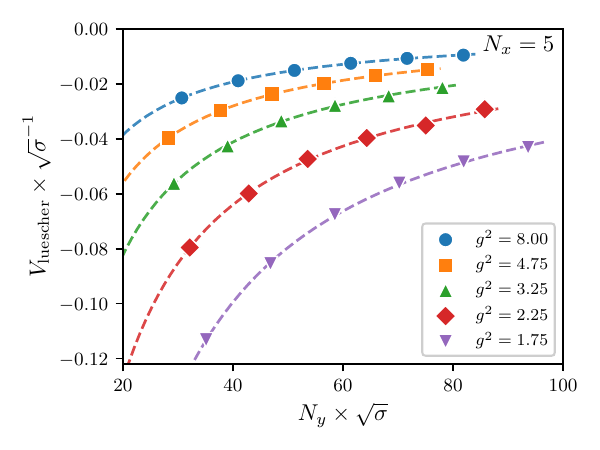}
  \includegraphics[width=0.95\columnwidth]{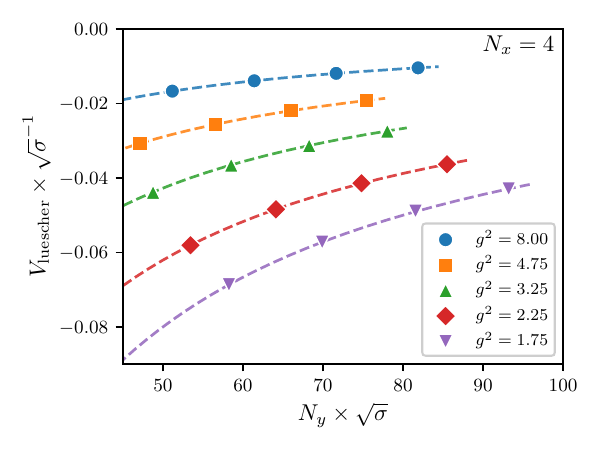}
  \includegraphics[width=0.95\columnwidth]{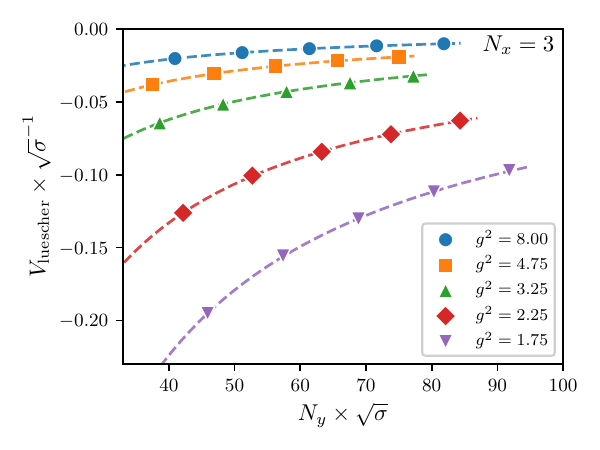}
  \caption{Lüscher term contribution of the potential for different $g^2$ in units of the string tension for the lattice widths $N_x$ of $5$ (top), $4$ (middle), and $3$ (bottom). $V_{\text{luescher}}=V(N_y)-\sigma N_y-c$.}
  \label{fig:1or}
\end{figure}

When fitting to the data of the potential $V(N_y)$, we also include
a term proportional to $\gamma/N_y$ corresponding to the Lüscher
term with $\gamma$ as a fit parameter. Indeed, the inclusion of this
term in the fit improves the  $\chi^2/\mathrm{dof}$ significantly. The
$1/r$ term indeed describes the data excellently, which can be seen in
\cref{fig:1or} where we plot $V(N_y) -\sigma N_y - c$ as a function of
$N_y$. The dashed line is the fitted curve, the shapes represent the 
data points. Note that the relative contribution of the Lüscher term
is still only around $0.1 \%$ for large $g^2$ and $0.5 \%$ for small
$g^2$, which is why the $1/r$ term is invisible in
\cref{fig:potfitexample}.

\begin{figure}
    \includegraphics[width=0.45\textwidth]{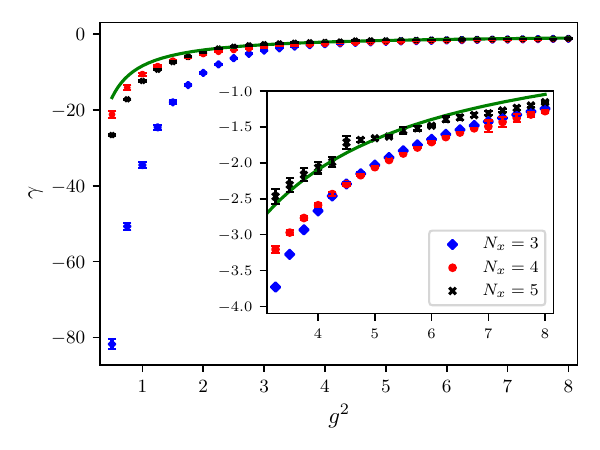}
    \includegraphics[width=0.45\textwidth]{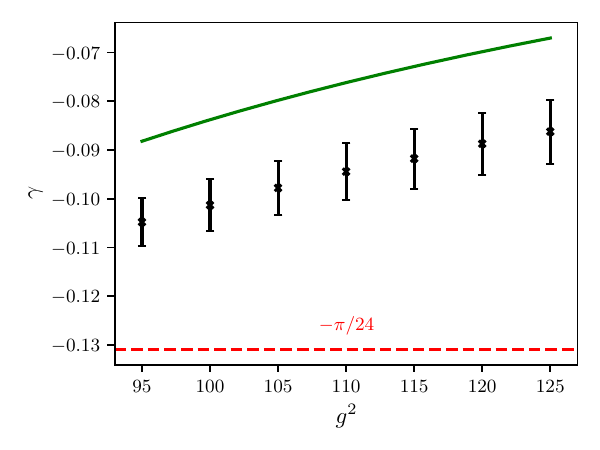}
    \caption{Best fit value of $\gamma$ as a function of $g^2$ for
      three different widths of the lattice $N_x=3$ (blue diamonds),
      $N_x=4$ (red dots) and $N_x=5$ (black crosses) with the
      prediction from the strong coupling expansion added in
      green and the expected continuum value of $-\pi/24$ in dashed red. Additional data points for $N_x=5$ at large coupling are
      added to confirm the scaling behaviour for the large coupling
      limit.}
    \label{fig:lusch}
\end{figure}

The results for the Lüscher term ($\gamma$ in \cref{eq:V}) are plotted
in \cref{fig:lusch} as a function of $g^2$. For $g^2<2$ the value of
$\gamma$ trends towards large negative values. 
For large $g^2$-values, on the other hand
the measured $\gamma$-values tend towards zero. While the
measured value of $\gamma$ is at $g^2=8$ still a factor of about eight 
larger in modulus than $-\pi/24$, the modulus is smaller 
than $\pi/24$ at $g^2=95$, see the lower panel.
Therefore, there is a $g^2$-value between $8$ and $95$
where $\gamma = -\pi/24$. Moreover, the observed value of $\gamma$
appears to approach zero slowly for $g^2\to\infty$.
At $g^2=\infty$, $\gamma$ must equal zero as the magnetic part in the
Hamiltonian becomes irrelevant, as also confirmed by the strong
coupling expansion.

In the upper panel of \cref{fig:lusch} we show the best fit value for $\gamma$ for three
different transversal widths $N_x$ of the lattice. At large values of
$g^2$, we observe a mild dependence on $N_x$, see the inset in the
figure. However, there appears to be no monotonic convergence in
$N_x$. $N_x=4$ results lie in between the $N_x=5$ and $N_x=3$
results only for a small intermediate region of about $3<g^2<4.5$. For
small values of $g^2$, finite size effects quickly become larger  
than $100\%$.

Finally, for the largest $N_x$-value, we observe a fish bone structure
in the $g^2$-value region between $4$ and $7$. This is a hysteresis-like effect introduced
by the DMRG on MPS algorithm used, which utilizes the converged state
of the next largest $g^2$ as the initial state for every new value of
$g^2$. It is usually too small to be relevant but can be seen here due
to the relatively small contribution of the Lüscher term in this model
that makes it more sensitive to such effects. The effect is
short-lived in the sense that, once a sufficient accumulation of error
has taken place, the DMRG algorithm will converge back to the correct
state. 

In both panels of \cref{fig:lusch} we also include the strong coupling
prediction for the Lüscher term as a solid line. At $g^2=100$ we still
observe a $15\%$ deviation of the expansion to the MPS
result. However, this could be a higher order effect in $1/g^2$ or
$1/R$. Also, finite volume effects cannot be excluded.

\subsection{String width $\omega^2$}

\begin{figure}
  \includegraphics[width=0.35\textwidth]{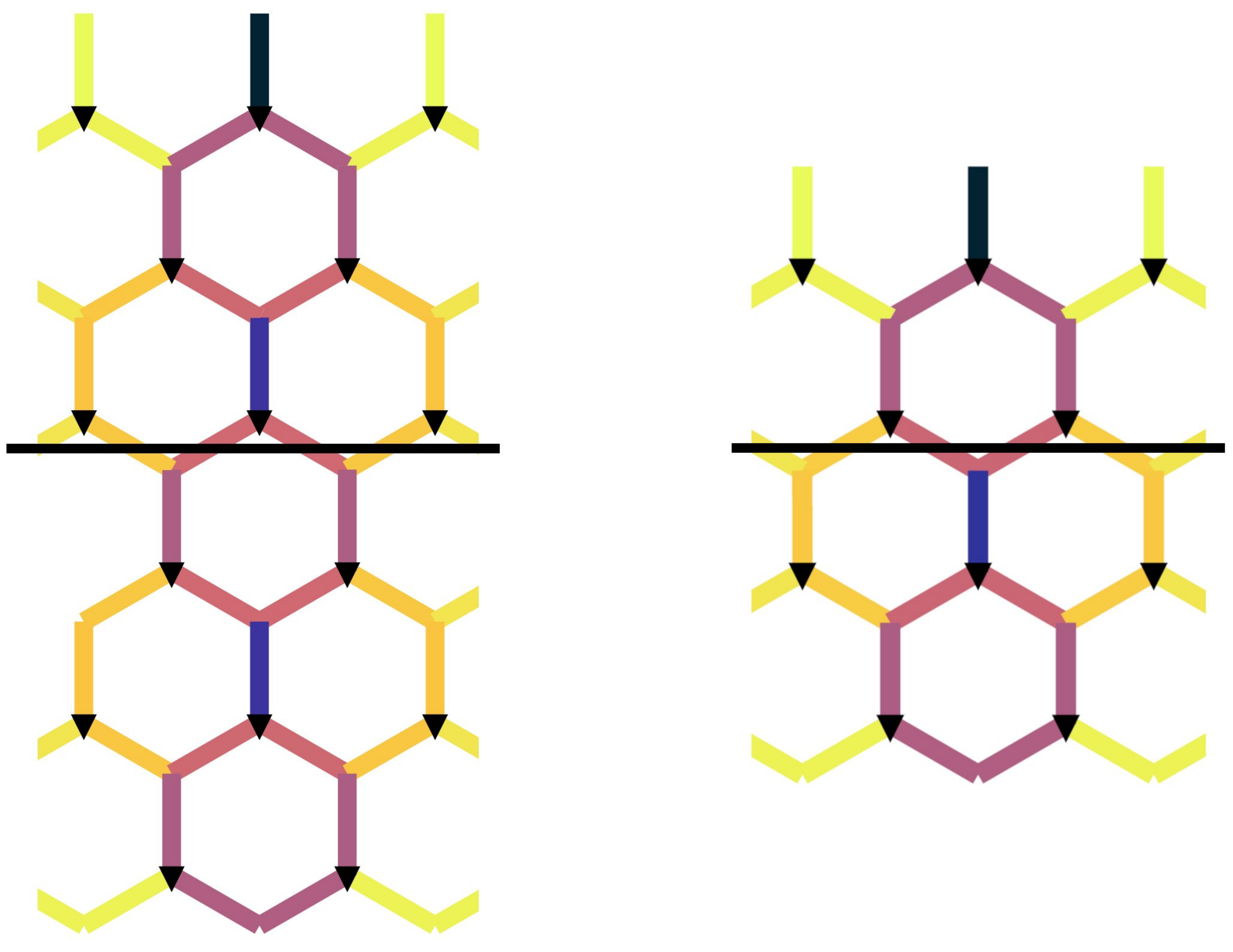}
  \caption{Exemplary plot of the transversal cut used to obtain the
    flux profile for $N_y=4$ (right) and $N_y=6$ (left) at $N_x=5$} 
  \label{fig:cut}
\end{figure}

\begin{figure}
    \includegraphics[width=0.45\textwidth]{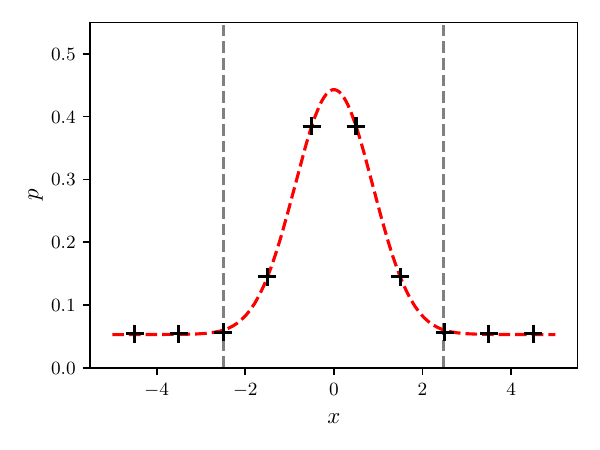}
    \includegraphics[width=0.45\textwidth]{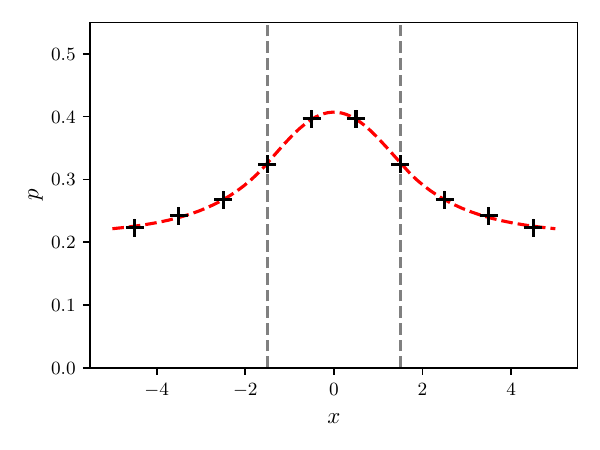}
    \caption{Flux distribution of $p = \langle \mathbb{1}-\ket{0} \bra{0}
      \rangle$ on each link centrally wrapped around the system over
      the orthogonal distance from the string centre $x$ for $g^2=8$
      (top) and $g^2=0.5$ (bottom) with $N_y=12$, fitted with
      \cref{eq:gaussian}. $x_0$ is plotted as grey dashed
      lines.}
    \label{fig:cross}
\end{figure}

As mentioned before, one property of rough strings is the logarithmic
scaling of its (squared) width $\omega^2$ with the length $r$.
$\omega^2$ can be extracted from the flux distribution of the
cross-section of the strings. For this we perform a cut through the
lattice in the transversal direction as close to the longitudinal
centre of the string as possible while maximising the number of cut
links at the same time. These criteria result in a cut slightly above
(below) the centre of the string, as visualised in \cref{fig:cut}.
The effect of this slight deviation from the longitudinal centre will
become negligible quickly with increasing string length.
Plotting the flux as a function of the distance from the transversal
centre as exemplarily shown in \cref{fig:cross}, reveals a Gaussian-like
distribution with heavy  
tails\footnote{Fitting with a pure Gaussian was attempted but failed
to capture the tails correctly for longer strings.} which is well-fitted 
by a Gaussian with smooth exponential \cite{Verzichelli:2025cqc} tails given by
\begin{align}
    \begin{split}
        \label{eq:gaussian}
        f(x) =
        \begin{cases}
        A \, \text{exp}\left({-\dfrac{x^2}{2\omega^2}}\right)+c & |x| \leq x_0 \\
        A \,
        \text{exp}\left({-\dfrac{x_0^2}{2\omega^2}-\dfrac{x_0}{\omega^2}
          \left(|x|-x_0\right)}\right)+c & |x| > x_0 
        \end{cases}
        .
    \end{split}
\end{align}
Here, $x_0$ represents the distance from the centre of the string from 
which on the heavy tails are added to the functional form. Thus,
$f(x)$ from \cref{eq:gaussian} depends on three fit parameters, $A,
\omega^2$ and $x_0$. Two exemplary fits with different couplings
$g^2=0.5$ and $g^2=8$ and string length $N_y=12$ are displayed in
\cref{fig:cross}. 

\begin{figure}
\includegraphics[width=0.45\textwidth]{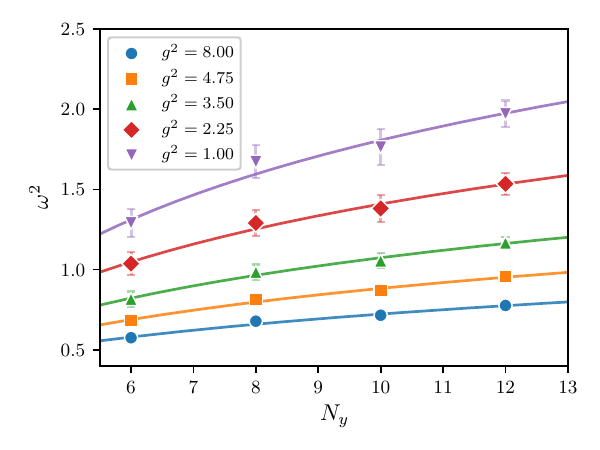}
    \caption{String widths $\omega^2$ over $N_y$ for selected $g^2$ and
      $N_x=5$. The solid lines represent fits to the data with
      logarithmic functional form \cref{eq:log}
      including all points with $N_y\geq6$ (see text) in the fit.}
    \label{fig:l2}
\end{figure}

By performing such fits for all couplings and lengths of the strings at the largest $N_x$ of $5$,
the scaling of the width $\omega^2$ with the length of the string can
be extracted. The results for $\omega^2$ are plotted for select $g^2$
values in \cref{fig:l2} as a function of $N_y\propto r$.

In general, the width increases with $N_y$. The observed
zigzag behaviour overlaying this increase originates again from the
geometry of the lattice: For $N_y$ divisible by $4$ the shortest path
for the string is via one single link in the longitudinal centre,
while for the other $N_y$ values there are two shortest paths with equal length.
This can be nicely seen in  \cref{fig:cut}.
Since this is not only the case at the centre of the string, this
effect becomes less and less relevant towards larger $N_y$ values.
In order to reduce the influence of this zigzag modulation, in the following we
 only consider $N_y\geq6$.

For a rough string, which ordinarily possesses a Lüscher term, one would
expect a logarithmic scaling of the width in the length of the
string. For rigid strings, on the other hand, the width is expected
to be constant as the length is increased. Clearly, our results for
$\omega^2$ are not well described by a constant, which is why we
attempt a fit of the form
\begin{equation}
    \label{eq:log}
    \omega^2(N_y) = A\ \text{ln}(N_y-b)
\end{equation}
to the data. As can be seen in \cref{fig:l2} for $N_y\geq6$ all data points are
well described by the fit.

\section{Discussion}
\label{sec:discussion}

The SU$(2)$ QLM investigated in this paper seems to exhibit
confinement for all values of $g^2$ and $N_x$ investigated. While this is
expected for this model, it appears remarkable to us how well the data
for the static potential fall on a universal curve once plotted in
appropriate units, cf.~\cref{fig:uni}. At least visually there are no
deviations from the linear behaviour observable up to values of
$N_y\sqrt{\sigma}>150$. For large $g^2$-values, the measured string
tension agrees well with the leading order strong coupling result.
Comparing $N_x=3,4$ and $N_x=5$, c.f. \cref{fig:tension}, we see finite size
effects in particular when comparing $N_x=3$ with $N_x=4$ and for
$g^2<6.5$. For all values of $g^2$ we investigated, the difference in
the string tension between $N_x=4$ and $N_x=5$ is already much smaller
than the difference between $N_x=3$ and $N_x=4$, indicating rapid
convergence in $N_x$. Thus, we expect no large finite size effects for
the results obtained with $N_x=5$. Moreover, the qualitative
dependence of $\sigma$ on $g^2$ remains unchanged for the three values
of $N_x$ we have available.

We find clear evidence for a Lüscher term in the potential with a
$g^2$ dependent coefficient. This is in qualitative agreement with the
strong coupling expansion. Like in the case of diagonal string in
Wilson's lattice gauge theory on a square lattice~\cite{Kogut:1981ny},
there is no roughening transition for the investigated strings in this
QLM, and $\gamma$ is $g^2$ dependent. However, unlike in
Ref.~\cite{Kogut:1981ny}, the universal value for $\gamma$ is not
approached due to a non-existent continuum limit.

From \cref{fig:lusch} finite size effects for $\gamma$ can be seen by
comparing $N_x=3, 4,$ and $N_x=5$. When comparing the three
$N_x$-values, we observe that for $g^2<2$ finite size effects are
large, and could even at $N_x=5$ be of the order of 100\%.
In the intermediate region of $g^2$-values the step from $N_x=4$ to
$N_x=5$ still amounts to a 20\% correction. In the strong coupling
region finite size corrections are milder, consistent with increasing
physical volumes with increasing $g^2$.
For $g^2>4$ or so the convergence in $N_x$ is also not monotonic.
Clearly, in the future larger values of $N_x$
need to be studied in order to reduce the corresponding uncertainty. 

Strong finite size effects on the Lüscher term were also observed in
Ref.~\cite{DiMarcantonio:2025cmf} in a $\mathbb{Z}_2$ Abelian gauge
theory. However, it should be mentioned that $N_x=5$ must be
multiplied by $\sqrt{3}\approx 1.7$ in comparison to
Ref.~\cite{DiMarcantonio:2025cmf}, and is thus in principle larger
than the maximal transversal extent of $6$ in lattice units used in
that reference, despite the fundamentally different theories.

Another point to highlight is that the error estimate for $\gamma$ is
not statistical but comes from the $\chi^2$ fit only. It is unclear to
us whether our error estimates are accurate such that the uncertainties
in \cref{fig:lusch} should be taken with care.

The Lüscher term has been studied previously using stochastic Monte
Carlo simulations in the Lagrangian
formulation~\cite{Bali:1994de,Luscher:2002qv} in particular also in
SU$(2)$~\cite{Caselle:2004er,HariDass:2007tx,Pepe:2009in}.  
In comparison, the major advantage of the Hamiltonian formalism is the
absence of statistical uncertainties, which grow in stochastic
simulations towards larger distances such that the signal-to-noise
ratio deteriorates exponentially, a problem which can be only solved
in principle by multi-level approaches~\cite{Luscher:2002qv,Barca:2024fpc}.
However, the results presented in this paper show that with the
current volumes feasible in TNS simulations this advantage does not
play out as yet.

The width of the string appears to scale
logarithmically with the length of the string for all values of the
coupling investigated here. While a linear dependence on the length of
the string is not excluded, it is certainly not independent of the
length of the string as expected for a rigid string. We interpret
this result as an indication for a rough string for all values of the
coupling investigated here. This
includes in particular also the values of the coupling where the
electric part in the Hamiltonian dominates the dynamics, in agreement
with the result from the strong coupling expansion.

\section{Conclusion \& Outlook}
\label{sec:conc}

In this paper we have investigated a $(2+1)$-dimensional non-Abelian
SU$(2)$ quantum link model using the $5$-dimensional
representation of the embedding group SO$(5)$. We have simulated this
theory with exact non-Abelian SU$(2)$ local gauge symmetry for the
first time using matrix product state algorithms on a hexagonal
lattice, which can be mapped particularly efficiently to the MPS. 

We have computed the static potential as a function of the distance by 
studying configurations with and without two separated static
charges. The first remarkable result is that the string tension $\sigma$ can be
extracted for all values of the coupling and transversal lattice extents to be positive, indicating a
confined theory for all values of the bare coupling. Using again the
string tension, we have then shown that there exists a unique potential in
units of $\sqrt{\sigma}$ for all values of the bare coupling. However,
the lattice spacing does not approach zero with $g^2\to0$, but
increases again for $g^2<4.687(2)$. Thus, as expected, a continuum
limit does not appear to exist for this QLM in the traditional sense.

Moreover, we have investigated the so-called Lüscher term, a $\gamma/r$
contribution to the potential at large distances, with
$\gamma=-\pi/24$ known from effective string theory in $2+1$
dimensions. However, as first discussed for diagonal strings in
Wilson's lattice gauge theories, due to the hexagonal lattice geometry
and the way the strings are placed on the lattice, there is no
roughening transition for the investigated strings and the Lüscher
term is $g^2$ dependent, in qualitative agreement with the strong
coupling expansion we have performed for this QLM.

Even if with the simulation volumes currently feasible in TN
simulations the Hamiltonian formalism is not yet competitive with
Monte Carlo, we expect that in the not too distant future it will be
possible to determine even subleading contributions to the potential
in the Hamiltonian formalism. This is mainly due to the absence of
statistical uncertainties, which on the other hand make it even more
important to control all the systematics.

Finally, the investigation of the width of the string as a function of
the string length indicates that the strings are rough again for all
values of the bare coupling. This conclusion is supported by the
squared string width increasing like the logarithm of the length of
the string in the whole range of $g^2$-values we investigated. This is
again in qualitative agreement with the strong coupling expansion.

In the future larger transversal lattice
extents or in general larger volumes must be investigated in order to
better control the associated finite-size effects. For the same reason
also other systematic effects of the method need to be thoroughly
studied in order to fully profit from the advantages of the
Hamiltonian formalism. It would be interesting to study other
geometries of the string for which a roughening transition can be expected.

Finally, given the results presented in this paper, it appears
promising and interesting to also investigate
larger representations of SO$(5)$ in order to understand QLMs
better in general, and to eventually prepare for simulations on
quantum devices.

\begin{acknowledgments}
  This project was funded by the Deutsche Forschungsgemeinschaft (DFG,
  German Research Foundation) as a project in the CRC 1639 NuMeriQS --
  project no.\ 511713970.
  The authors gratefully acknowledge the access to the Marvin cluster
  hosted and provided by the University of Bonn.
  We thank Uwe-Jens Wiese for helpful and illuminating discussions on
  QLMs and this project, and for constructive comments to a first version of this manuscript.
  We thank all members of the Bonn lattice QCD groups for the most enjoyable collaboration.
\end{acknowledgments}

\bibliography{bibliography}

\end{document}